\documentclass{eas}
\usepackage{graphicx}
\def\dg{^{\circ}}

\def\simlt{\mathrel{\spose{\lower 3pt\hbox{$\mathchar"218$}}\raise 2.0pt\hbox{$\mathchar"13C$}}}
\def\simgt{\mathrel{\spose{\lower 3pt\hbox{$\mathchar"218$}}\raise 2.0pt\hbox{$\mathchar"13E$}}}
\begin{document}

\title{PILOT and cosmic shear}
\runningtitle{PILOT and cosmic shear}

\author{W. Saunders}\address{Anglo-Australian Observatory, Epping, NSW 1710, Australia; \email{will@aao.gov.au}}
\begin{abstract}

Cosmic shear offers a remarkbly clean way to measure the equation of state of the Universe and its evolution. Resolution over a wide field is paramount, and Antarctica offers unique possibilities in this respect. There is an order of magnitude gain in speed over temperate sites, or a factor three in surface density. This means that PILOT outperforms much larger telescopes elsewhere, and can compete with the proposed DUNE space mission.

\end{abstract}
\maketitle

\section{Introduction}

Weak lensing has been widely identified as the most promising route to measuring the evolution of the equation of state of the Universe, and hence to understanding the nature of dark energy. A large number of huge lensing surveys are planned: Panstarrs, LSST, The Dark Enery Survey on the CTIO 4-m, KIDS on VST, HyperSuprimeCam for Subaru; and the DUNE and SNAP satellites.

For weak lensing, resolution is paramount, because the galaxies used to measure the effect are $<1''$ in size, and must be at least partially resolved. Antarctica offers unique possibilities for wide-field, high-resolution imaging, so even modest sized telescopes may offer world-beating performance. This paper is a preliminary investigation into the possibilities for measuring weak lensing with PILOT - the Pathfinder for an International Large Optical Telescope, a 2-m class optical/infrared telescope proposed for Dome C on the Antarctic Plateau with first light at the end of 2012.

We assume a 2.4m f/10 telescope, with a $0.75deg^2$ CCD camera, with image scale $\sim 0.1''$/pixel.

\section{The size of lensed galaxies and effect of resolution on sensitivity}

Figure 1 below shows the size of galaxies detected in the Hubble Deep Field, as a function of AB magnitude. For $AB=25-27^m$, most galaxies are between $0.2''$ and $0.4''$, with weak dependence on magnitude. This means temperate-site observations, with image quality $ 0.6''$ or more, struggle to determine ellipticities for these galaxies.

The integration time required to measure ellipticity is a dramatic function of image quality, because (a) the intrinsic ellipticity of the galaxy is diluted by the PSF, and it must be observed to higher S/N to compensate, and (b) the observations are sky-limited, and the overall sky noise increases with the observed image size.

\begin{figure}
\begin{center}
\includegraphics[width=9.5cm,angle=-90]{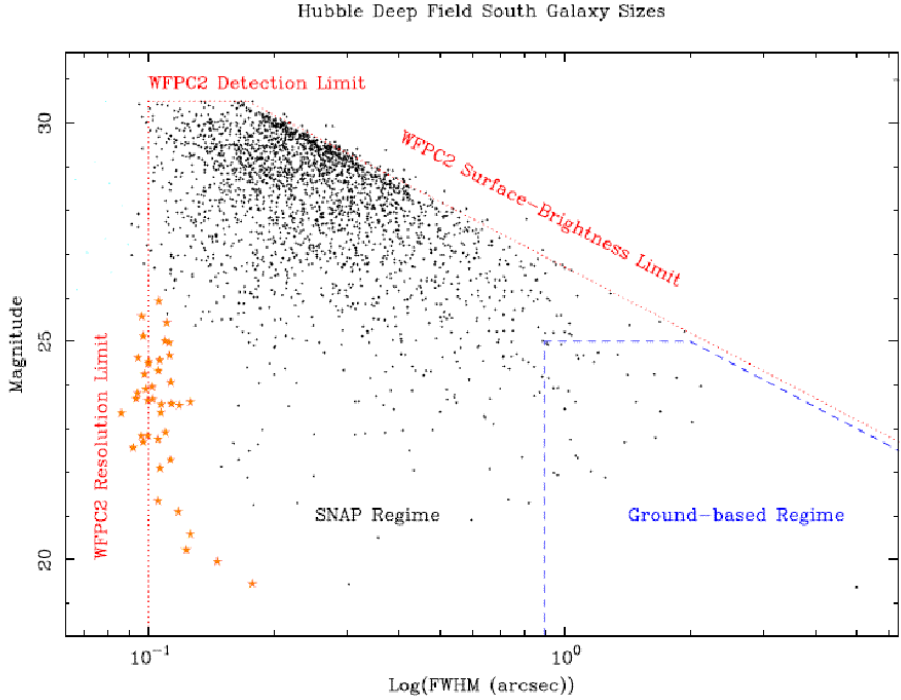}
\end{center}
\caption{Image size for galaxies from the HDF, as a function of AB magnitude. Taken from Curtis {\em et al.\/} 2000.}
\end{figure}

Suppose we are trying to measure the ellipticity of a galaxy of intrinsic FWHM  $a \times b$, with observed dimensions (including seeing, telescope optics, pixellation) $A \times B$. Assuming Gaussian statistics, the time required to measure the ellipticity of a galaxy to a given S/N scales as \\

$t_{req} \propto A B (A^2+B^2)(A+B)^2 /(a^2-b^2)$ \\

For a given intrinsic galaxy shape and size, $t_{req}$ then varies as the {\em sixth power} of the observed image size. Figure 2 shows this effect of image size and seeing on the required integration time for fixed aperture, object brightness and background level. Galaxies smaller than the seeing are very difficult to measure because the apparent ellipticity is so small; large galaxies are difficult because the sky noise is large. The optimal size for target galaxies is 1-3 times the seeing, which means Dome C is ideally suited for this work. For typical galaxies, the required integration times at Dome C are one to two orders of magnitudes less than for a similar telescope at Mauna Kea.

\begin{figure}
\begin{center}
\includegraphics[width=12.5cm]{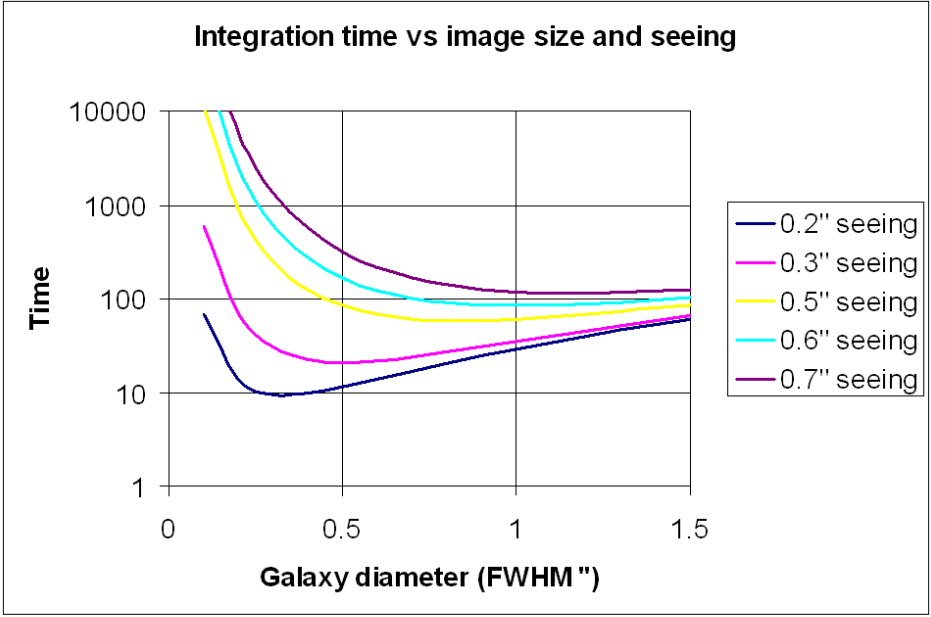}
\end{center}
\vspace{-0.75cm}
\caption{The overall survey speed as a function of seeing and galaxy size.}
\end{figure}
\begin{figure}
\begin{center}
\includegraphics[width=12.5cm]{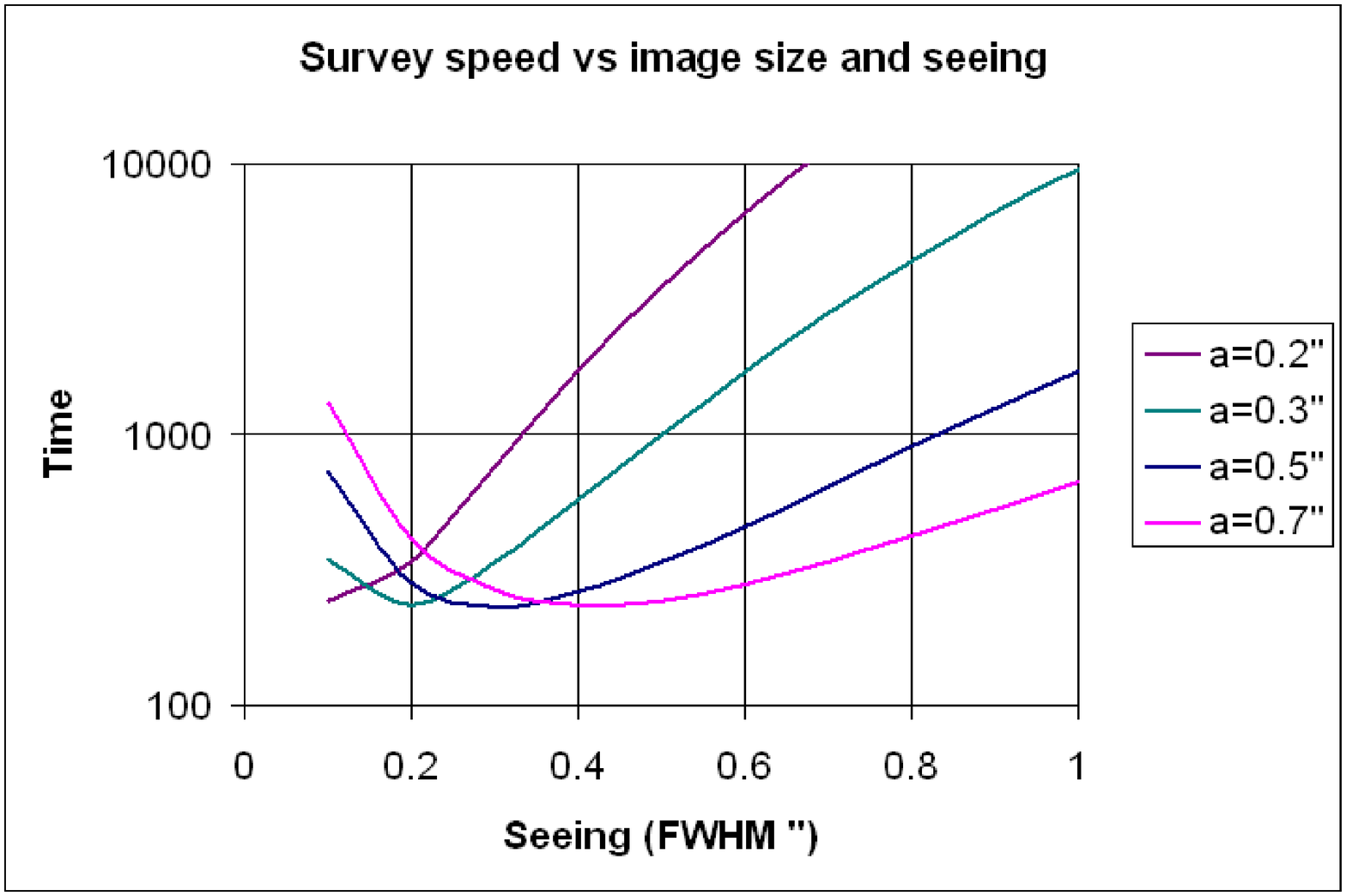}
\end{center}
\vspace{-0.75cm}
\caption{Required integration time per galaxy as a function of seeing and galaxy size.}
\end{figure}

Of course, if we match the pixel scale to the median seeing, then worse seeing means we can image more sky at once for given detector area. Figure 3 shows the overall lensing survey speed as a function of image quality and galaxy size, for a telescope of fixed aperture and detector size, but with pixel scale matched to the seeing. The gain in overall survey speed between Dome C and the best temperate sites is still an order of magnitude.

\section{Image quality with PILOT}

The estimated PILOT wide field image quality, including median seeing, has been presented in Saunders {\em et al.\/} 2008. For $i$-band, the value is $0.3''$. This includes the effect of a fast tip-tilt secondary, which corrects for residual low-level turbulence above the telescope.

If we could correct for high level turbulence, we could improve the image quality much further. The angle over which the high level turbulence is correlated is much smaller than the field, so we have two requirements: a density of guide stars high enough to sample this turbulence, and some way of making differential tip-tilt corrections across the field. The latter could be achieved via a deformable mirror, but it could also make use of Orthogonal Transfer CCDs (Tonry, Burke and Schecter 1997), which allow charge shuffling in both directions, in $\sim 500 \times 500$ pixel sub-arrays.

The median Fried length $r_0$ for PILOT at $i$-band is about 65cm, so $D/r_0 \sim 4$. This is exactly the regime where tip-tilt correction provides the largest gains (Hardy 1998, Jenkins 2000). The image quality that can be expected in this regime has been comprehensively explored by Jenkins (1998) and also by Kaiser, Tonry and Luppino (2000, henceforth KTL). KTL conclude that at Mauna Kea, the density of suitable guide stars is too low to map the required tip-tilt corrections over wide fields. However, at Dome C we have three large gains over Mauna Kea: (a) we can have a 2.5 times larger collecting area while preserving $D/r_0$=4, (b) the Greenwood timescale is 3 times longer, and (c) the isoplanatic angle is 2.5 times larger (Agabi {\em et al.\/} 2006, Lawrence {\em et al.\/} 2004). So we can use fainter guide stars, and we can also tolerate larger separations between them. In overall terms of (guide stars)/(isoplanatic angle)$^2$, we are 20 times better off than Mauna Kea. This is enough to completely map the deflection field even at high latitudes, leaving negligible residual isoplanatic error (KTL figure 15).

The tip-tilt correction gives a $\frac{1}{3}$ improvement in the encircled energy diameter. It also gives a significant diffraction-limited core containing 28\% of the energy, with a Strehl ratio of $\sim 0.25$ (KTL, Jenkins 1998). 

So we can expect diffraction-limited cores, with FWHM $0.07''$, across the whole field. The limiting factor in practice will be the pixel size ($0.075- 0.1''$), and the limited ability of OTCCDs to do sub-pixel shuffling. The gain in encircled energy can be fully utilised, to give us a $FWHM \sim 0.2''$. This is perfectly matched both to our target galaxies, and to the pixel scale for the telescope (which is determined also by NIR considerations). It is better than the proposed DUNE satellite ($0.23''$).

\section{Comparison with other projects}

For surveying ellipticities of typical faint galaxies, PILOT is an order of magnitude more efficient, for given telescope and detector areas, than telescopes at temperate sites. In practice, this means that we can reach surface densities not realistic elsewhere. It's very hard to reach densities higher than 10-20 galaxies$/arcmin^2$ even with 8-m class telescopes, compared with 50-100 from space (e.g. Kasliwal {\em et al.\/}). For PILOT, we should reach S/N = 10 for $i_{AB}=26^m$, $0.3''\times 0.2''$ FWHM galaxies in $5000-10,000s$, giving us $\sim 50$ galaxies$/arcmin^2$. A high space density is crucial for measuring the peak of the lensing power spectrum at $l \sim 5000-10,000$ ($2-4'$ scales), which is where the greatest sensitivity to cosmological model occurs (Figure 4a). The higher resolution also means that the median depth of the lensed galaxies is greater, which increases (a) the sensitivity to lensing, (b) the volume surveyed, and (c) the lever-arm to measuring the evolution of the power spectrum.

\begin{figure}[h]
\begin{center}
\includegraphics[width=12.5cm]{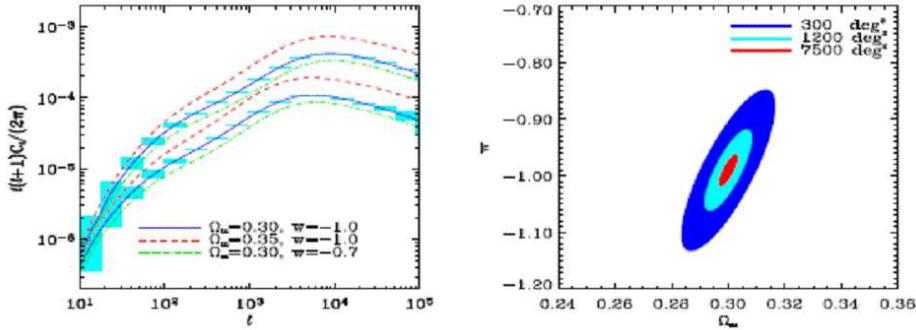}
\end{center}
\caption{(a) Power spectrum and error bars for a $1200 deg^2$ weak lensing survey. (b) Cosmological constraints for surveys of different sizes. The survey proposed for PILOT is $2500 deg^2$. Both plots traken from the proposal to the ESA Cosmic Vision program.}
\end{figure}

Because the effects looked for are so subtle, a fundamental limitation with ground-based imaging surveys is the constancy of the PSF across the field, and with time, elevation, colour, etc. PILOT has two large advantages over temperate telescopes: (a) the lensing signal is less diluted by resolution, so the sensitivity to systematic error is less by a factor of several, and (b) the PILOT optics deliver superb imaging, because of the slow f-ratio (f/10) and modest field of-view ($1\dg$), and because we have a Ritchey-Chr\'etian design. The optical design is diffraction-limited at $i$-band across the entire field (Saunders {\em et al.\/} 2008).

There is also excellent overlap with the Dark Energy Survey, which will obtain photometry in $griz$ bands, potentially giving photometric redshifts for the entire survey, and also the South Pole Telescope, allowing independent and robust determination of cluster masses, as a separate probe of the evolution of the power spectrum.

Compared with the DUNE satellite, a PILOT lensing survey looks remarkably good. The overall survey speeds, taking into account the image quality, sky background, aperture and field of view, are about the same. PILOT can only access $\sim \frac{1}{4}$ of the high latitude sky. PILOT will not be capable of getting the deep Y, J, H-band data proposed for DUNE, but will be capable of obtaining deep $g,r$ and also $K_{dark}$ data if suitable data is not available elsewhere.

If 50\% of dark and grey time is allocated to a lensing survey, PILOT could survey all accessible high latitude sky (about $5000deg^2$) to $i_{AB}=25^m$ in 4 years. On the (untested) assumption that adequate photometric redshifts can be obtained, then the overall signal-to-noise for cosmological parameter estimation is similar to the DUNE weak lensing survey (Figure 4b), but the survey would be completed by the time DUNE is launched.


\end{document}